\documentclass[showpacs,10pt,twocolumn,prl]{revtex4-1}

\usepackage{amsmath}
\usepackage{amssymb}
\usepackage{graphicx}
\usepackage{amssymb}
\usepackage{graphics}
\usepackage{epsfig}
\usepackage{CJK}
\usepackage{color}
\usepackage{soul}

\begin{document}

\begin{CJK*}{GBK}{Song}
\title{Thermoelectricity and electronic correlations enhancement in FeS by slight Se substitution}
\author{Yu Liu,$^{1,\dag,\ddag,*}$ Aifeng Wang,$^{1,\|,\S,*}$ V. N. Ivanovski,$^{2}$ Qianheng Du,$^{1,3,\sharp}$ V. Koteski,$^{2}$ and C. Petrovic$^{1,3,\P}$}
\affiliation{$^{1}$Condensed Matter Physics and Materials Science Department, Brookhaven National Laboratory, Upton, New York 11973, USA\\
$^{2}$Department of Nuclear and Plasma physics, Vinca Institute of Nuclear Sciences - National Institute of the Republic of Serbia, University of Belgrade, Belgrade, Serbia\\
$^{3}$Department of Materials Science and Chemical Engineering, Stony Brook University, Stony Brook, New York 11790, USA}
\date{\today}

\begin{abstract}
  We report thermoelectric studies of FeS$_{1-x}$Se$_x$ ($x$ = 0, 0.06) superconducting single crystals that feature high irreversibility fields and critical current density $J_c$ comparable to materials with much higher superconducting critical temperatures ($T_c$'s). The ratio of $T_c$ to the Fermi temperature $T_F$ is very small indicating weak electronic correlations. With a slight selenium substitution on sulfur site in FeS both $T_c$/$T_F$ and the effective mass $m^*$ rise considerably, implying increase in electronic correlation of the bulk conducting states. The first-principle calculations show rise of the density of states at the Fermi level in FeS$_{0.94}$Se$_{0.06}$ when compared to FeS which is related not only to Fe but also to chalcogen-derived electronic states.
\end{abstract}
\maketitle
\end{CJK*}

\section{INTRODUCTION}

Since the discovery of La(O$_{1-x}$F$_x$)FeAs ($x$ = 0.05-0.12) with a superconducting transition temperature $T_c$ = 26 K \cite{Hosono}, the layered iron-based superconductors have been extensively studied. Stoichiometric tetragonal FeS showing a $T_c$ below 5 K was synthesized in 2015 by using a hydrothermal method \cite{Lai}. It shares the same simplest PbO-type structure with FeSe \cite{FeSe}, which motivated many subsequent studies devoted to investigation of its superconducting state and relation to crystal structure details.

In contrast to FeSe, the X-ray diffraction experiment indicated no structural transition down to 10 K in FeS \cite{Pachmayr,Kuhn}. The $T_c$ of FeS continuously decreases for hydrostatic pressures up to 2.2 GPa \cite{Holenstein}, while a second superconducting dome for higher pressures from 5 to 22.3 GPa has been observed with an enhanced $T_c$ \cite{Zhang}. The neutron scattering experiment and angle-resolved photoemission spectroscopy (ARPES) measurement indicated that FeS is less correlated when compared to FeSe \cite{Man,Reiss}, however the scanning tunneling microscopy (STM) measurement suggested a strong-coupling superconductivity with a large gap ratio of $\sim$ 4.65 in FeS \cite{Yang1}. The thermal conductivity and heat capacity measurements suggested a nodal or highly anisotropic superconducting gap structure in FeS \cite{Ying,Xing}, while a $\mu$SR study proposed that a full-gap state coexists with low-moment disordered magnetism \cite{Holenstein}. A large upper critical field anisotropy {$H_{c2}^{\parallel ab}$(0)}/$H_{c2}^{\parallel c}$(0) $=$ 10 has been observed in angle-dependent magnetoresistance and quantum oscillations, in line with the band structure calculation \cite{Borg,Terashima,Lin,Yang2}. The nonlinear behavior of Hall resistivity indicated multiband conductivity \cite{Lin}. Interestingly, the critical current density $J_c$ in FeS is comparable to other iron-based superconductors with much higher $T_c$, and it can be enhanced three times by 6\% Se doping \cite{AF1,AF2}, pointing that FeS-based materials with higher $T_c$ are promising for high-magnetic-field applications.

Thermopower is an effective parameter to characterize the nature and sign of carries as well
as the correlation strength in superconductors \cite{Sefat,Kang,Mun,Pourret,Matusiak,Butch,Kefeng1,Kefeng2,Collignon}. Fundamentally, the thermopower is entropy per carrier. To evaluate the correlation strength of electronic states in the bulk that may influence anisotropy of both normal state and the $J_c$ as well as the current-carrying capability in grain boundaries \cite{AndersenBM,WolfFA,YoYJ,HecherJ,YuSL,SkornyakovSL}, in this work we perform thermal transport studies of superconducting FeS$_{1-x}$Se$_x$ ($x$ = 0 and 0.06) single crystals. The Fermi temperature $T_F$ in both crystals is about an order of magnitude smaller when compared to common values in metals. Rather small $T_c$/$T_F$ suggests weak electronic correlations. Yet, our combined experimental and theoretical analysis indicates an increase of electronic correlations for a change in $x$ as small as 0.06, i.e. in FeS$_{0.94}$Se$_{0.06}$ when compared to FeS.

\section{EXPERIMENTAL DETAILS}

Single crystals of FeS$_{1-x}$Se$_x$ ($x$ = 0, 0.06) were synthesized by a hydrothermal method and characterized as described previously \cite{AF1,AF2}. X-ray diffraction (XRD) data were acquired on a Rigaku Miniflex powder diffractometer with Cu $K_{\alpha}$ ($\lambda=0.15418$ nm). The in-plane resistivity $\rho$ and thermopower $S$ were measured in a Quantum Design PPMS-9 system with standard four-probe technique. Thermopower was measured by using one-heater-two-thermometer setup with hooked copper leads using silver paint contact directly on crystals with typical dimensions $\sim$ 4$\times$3$\times$0.3 mm$^3$. Continuous measuring mode was adopted for thermopower measurement with the maximum heater power and period set as 50 mW and 1430 s along with the maximum temperature rise of 3$\%$. The relative error in our measurement for thermopower was below 5\% based on Ni standard measured under identical conditions. The Hall resistivity $\rho_{xy}$ was measured using standard four-probe method with the current flowing in the $ab$ plane and the magnetic field along the $c$ axis. In order to effectively eliminate the longitudinal resistivity contribution due to voltage probe misalignment, the Hall resistivity was obtained by the difference of transverse resistance measured at positive and negative fields, i.e. $\rho_{xy} = [\rho(+\mu_0H)-\rho(-\mu_0H)]/2$. The sample dimensions were measured by an optical microscope Nikon SMZ-800 with 10 $\mu$m resolution. The heat capacity was measured on warming procedure between 0.35 and 7 K by the heat pulse relaxation method in a Quantum Design PPMS-9.

The density of states of both FeS and Fe$_{16}$S$_{15}$Se$_1$ were calculated by the VASP code in the gradient corrected (GGA) implementation \cite{PhysRevB.54.11169} using the experimental value of FeS lattice parameter $c$. Lattice parameter $a$ and internal parameter $u$ were DFT-relaxed. The non-magnetic ground states were obtained via the Methfessel-Paxton order 1 integration method of the Brillouin zone using a grid of 13 $\times$ 13 $\times$ 9 $k$-points. The maximum forces are less than 0.02 eV \AA$^{-1}$ and the energy criterion was set to 10$^{-5}$ eV.

\section{RESULTS AND DISCUSSIONS}

Figure 1(a) shows the single crystal XRD patterns of tetragonal FeS$_{1-x}$Se$_x$ ($x$ = 0 and 0.06). Only (00l) reflections were observed for both crystals, yielding the $c$-axis lattice parameter $c$ = 5.035(5) and 5.061(22) {\AA} for FeS and FeS$_{0.94}$Se$_{0.06}$, respectively. The crystal structure as presented in the inset in Fig. 1(a); it is composed of a stack of edge-sharing FeS$_4$ tetrahedra layer by layer, without a spacer layer \cite{Lai}. The FeS$_4$ tetrahedron is nearly regular with two S-Fe-S angles very close to the ideal value of 109.5$^\circ$ \cite{Lee}.

The in-plane resistivity $\rho$(T) for FeS$_{1-x}$Se$_x$ ($x$ = 0 and 0.06) [Fig. 1(b)] monotonically decreases with decreasing temperature showing good metallic conductivity. The residual resistivity ratio $RRR$ = $\rho$(300 K)/$\rho$(5 K) is $\sim$ 68 for FeS, indicating low defect scattering. The value of $RRR$ decreases to $\sim$ 27 for FeS$_{0.94}$Se$_{0.06}$ with S/Se substitution disorder. The $\rho$(T) of FeS$_{1-x}$Se$_x$ ($x$ = 0 and 0.06) shows a positive curvature all the way up to room temperature, in contrast to FeSe where a negative curvature is generally observed at high temperatures \cite{FeSe}. An abrupt resistivity drop can be clearly seen [Fig. 1(c)], signaling the onset of superconductivity. Zero resistivity is observed at $T_c$ = 4.5(1) K for FeS, in agreement with the previous reports \cite{Lin}. The $T_c$ is slightly lower with a value of 4.1(1) K for FeS$_{0.94}$Se$_{0.06}$.

\begin{figure}
\centerline{\includegraphics[scale=1]{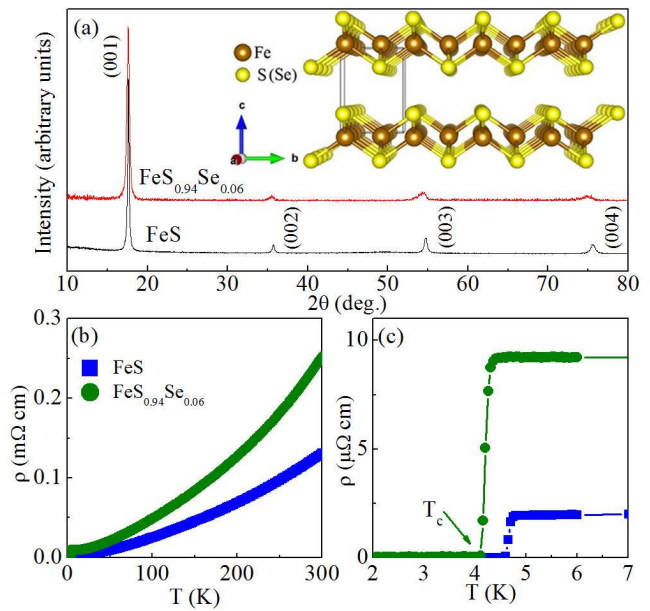}}
\caption{(Color online) (a) Single crystal X-ray diffraction (XRD) patterns of FeS$_{1-x}$Se$_x$ ($x$ = 0, 0.06). Inset shows the crystal structure with the space group $P4/nnm$. (b) Temperature dependence of in-plane resistivity $\rho$(T) for FeS$_{1-x}$Se$_x$ ($x$ = 0, 0.06) single crystals. (c) The enlargement of superconducting transitions at low temperature.}
\label{XRD}
\end{figure}

\begin{figure}
\centerline{\includegraphics[scale=1]{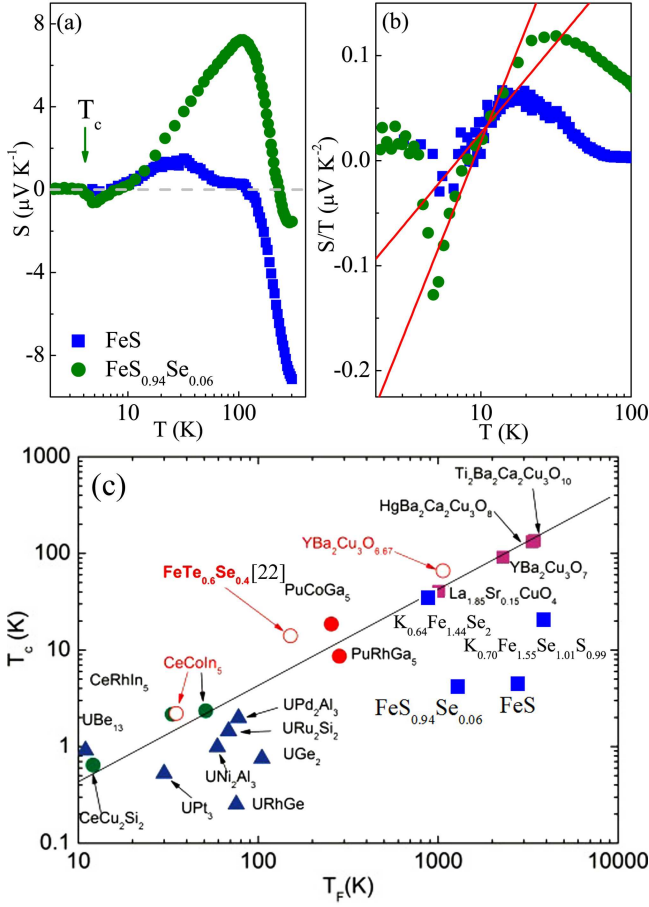}}
\caption{(Color online) Temperature dependence of in-plane (a) thermopower $S$(T) and (b) $S$/$T$ for FeS$_{1-x}$Se$_x$ ($x$ = 0 and 0.06) single crystals. (c) The Moriya-Ueda plot: $T_c$ as a function of $T_F$ for indicated superconductors \cite{Pourret,Sarrao,Moriya}; Four blue solid squares represent K$_{0.64}$Fe$_{1.44}$Se$_2$, K$_{0.70}$Fe$_{1.55}$Se$_{1.01}$S$_{0.99}$ \cite{Kefeng2}, and FeS$_{1-x}$Se$_x$ ($x$ = 0, 0.06).}
\label{MTH}
\end{figure}

Figure 2(a) shows the temperature dependence of in-plane thermopower $S$(T) for FeS$_{1-x}$Se$_x$ ($x$ = 0 and 0.06). At room temperature, the $S$(T) for both samples shows a negative value, consistent with dominant negative charge carriers \cite{Lin}. It is interesting that the value of $S$(T) decreases with decreasing temperature and becomes positive at $\sim$ 122(4) and 228(4) K for FeS and FeS$_{0.94}$Se$_{0.06}$, respectively. For pure FeS, the positive $S$(T) gradually increases with a much smaller slope and features a broad peak [$\sim$ 1.3(2) $\mu$V K$^{-1}$] around 28(9) K, then decreases again as temperature is lowered. The sign change of $S$(T) implies multiband transport in FeS, in agreement with the nonlinear Hall resistivity analysis \cite{Lin}. Even though the Hall coefficient $R_H$ is unchanged in this temperature range \cite{Lin}, the sign of $S$(T) might change due to a different dependence on the carrier density $n_e$ ($n_h$), mobility $\mu_e$ ($\mu_h$), and $S_e$ ($S_h$) in the two-band model \cite{Smith}:
\begin{equation}
R_H = \frac{1}{e}\frac{n_h\mu_h^2-n_e\mu_e^2}{(n_h\mu_h+n_e\mu_e)^2},
\end{equation}
\begin{equation}
S = \frac{S_en_e\mu_e+S_hn_h\mu_h}{n_e\mu_e+n_h\mu_h}.
\end{equation}
The positive $S$(T) of FeS$_{0.94}$Se$_{0.06}$ changes its slope from negative to positive temperature dependence at 107(3) K with a larger maximum value of 7.22(1) $\mu$V K$^{-1}$. With further decrease in temperature, the S(T) for both samples becomes negative again and vanishes at $T_c$ since the Cooper pairs carry no entropy. The $S$(T) of FeSe is positive at room temperature of $\sim$ 10 $\mu$V K$^{-1}$, then decreases with decreasing temperature and changes its sign to be negative below 200 K and back to be positive again near 30 K above $T_c$, featuring a minimum value of -10 $\sim$ -15 $\mu$V K$^{-1}$ around 115 K \cite{YJSong,Caglieris,Pallecchi}. The similar multiple sign changes in $S$(T) suggests that almost compensated electrons and holes complete in FeS$_{1-x}$Se$_x$. From the band structure
calculations \cite{Subedi}, indeed, both the hole and electron pockets appear in the identical Fermi-surface area. However, the complex details of $S$(T) make it difficult to be reproduced by the theory calculation using constant relaxation time approximation \cite{Caglieris}, calling for further improved method taken into consideration the correlation effects in this system.

\begin{figure}
\centerline{\includegraphics[scale=1]{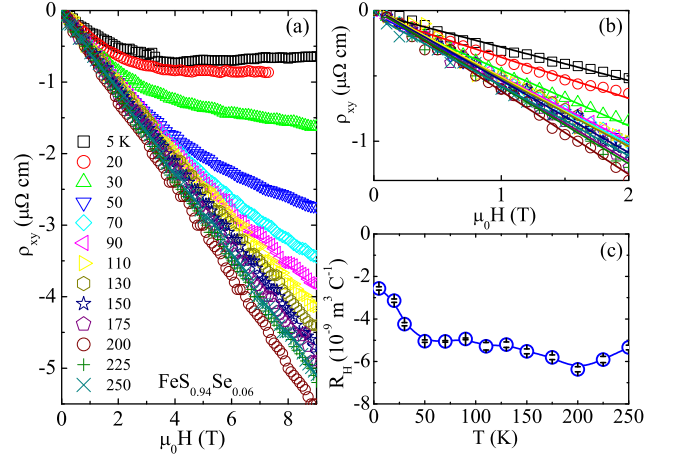}}
\caption{(a) Field dependence of the Hall resistivity $\rho_{xy}$ for FeS$_{0.94}$Se$_{0.06}$ at various temperatures. (b) Plots of $\rho_{xy}$ vs $\mu_0H$ below 2 T with solid linear fitting curves. (c) Temperature dependence of Hall coefficient $R_H$ derived from linear fit in (b) below 2 T.}
\end{figure}

Generally, the phonon drag $S_{\textrm{drag}}$ related to electron-phonon coupling gives $\propto T^3$ dependence for $T \ll \Theta_\textrm{D}$, $\propto T^{-1}$ for $T \gg \Theta_\textrm{D}$, and a peak structure at $\sim \Theta_\textrm{D}/5$, where $\Theta_\textrm{D}$ is the Debye temperature \cite{Barnard,Cohn}. The derived $\Theta_\textrm{D}$ is about 230 K for FeS \cite{Xing}. Since there is no peak structure at $\Theta_\textrm{D}/5$ = 50.7(1) K for FeS and thermopower is linear for FeS$_{0.94}$Se$_{0.06}$ around 49.1(1) K (see below for the values of $\Theta_\textrm{D}$), it is unlikely that the sign and slope change in $S$(T) is related to the phonon drag effect.

Figure 2(b) shows the temperature dependent of $S$(T) divided by temperature for FeS$_{1-x}$Se$_x$ ($x$ = 0 and 0.06). The zero-temperature extrapolated value of $S/T$ increases with Se doping; it is $\sim$ -0.15(1) $\mu$V K$^{-2}$ for FeS and reaches -0.34(3) $\mu$V K$^{-2}$ for FeS$_{0.94}$Se$_{0.06}$, with a magnitude proportional to the strength of electron correlation \cite{Behnia}. The $S$ is usually given by \cite{Barnard,Miyake,Behnia},
\begin{equation}
\frac{S}{T} = \pm \frac{\pi^2}{2}\frac{k_\textrm{B}}{e}\frac{1}{T_\textrm{F}} = \pm\frac{\pi^2}{3}\frac{k_\textrm{B}^2}{e}\frac{N(\varepsilon_\textrm{F})}{n},
\end{equation}
where $e$ is the electron charge, $k_\textrm{B}$ is the Boltzmann constant, $T_\textrm{F}$ is the Fermi temperature, which is related to the Fermi energy $\varepsilon_\textrm{F}$ and the density of states $N(\varepsilon_\textrm{F})$ as $N(\varepsilon_\textrm{F})$ = $3n/2\varepsilon_\textrm{F}$ = $3n/k_\textrm{B}T_\textrm{F}$, and $n$ is the carrier concentration (the positive sign is for hole and the negative sign is for electron). In a multiband system, it gives the upper limit of $T_\textrm{F}$ of the dominant band. Therefore we can extract $T_\textrm{F}$ = $2.8(2)\times10^3$ K for FeS and $1.25(11)\times10^3$ K for FeS$_{0.94}$Se$_{0.06}$, respectively. The ratio of $T_\textrm{c}/T_\textrm{F}$ characterizes the correlation strength in superconductors. For example, the $T_\textrm{c}/T_\textrm{F}$ is close to 0.1 in Fe$_{1+y}$Te$_{1-x}$Se$_x$ \cite{Pourret}, pointing to the importance of electronic correlation, while it is $\sim$ 0.02 in BCS superconductor, such as LuNi$_2$B$_2$C \cite{Pourret}. The value of $T_\textrm{c}/T_\textrm{F}$ is $\sim$ 0.0016(1) for FeS and $\sim$ 0.0033(2) for FeS$_{0.94}$Se$_{0.06}$, indicating weak electronic correlations but also a substantial enhancement for rather small amount of Se substitution on S atomic site. The value of $T_\textrm{c}/T_\textrm{F}$ is also smaller than that of K$_x$Fe$_{2-y}$Se$_2$ $\sim$ 0.04 \cite{Kefeng1}, but comparable with that of K$_x$Fe$_{2-y}$Se$_{1-z}$S$_z$ $\sim$ 0.005 \cite{Kefeng2}, which are also added into the Moriya-Ueda plot for comparison [Fig. 2(c)] \cite{Pourret,Sarrao,Moriya}.

Figure 3 shows the Hall resistivity $\rho_{xy}$ of FeS$_{0.94}$Se$_{0.06}$. At low temperatures, the $\rho_{xy}$ shows a nonlinear correlation with the magnetic field, similar to FeS \cite{Lin}, confirming a multiband effect. With increasing temperature, the nonlinear $\rho_{xy}$ gradually evolves to linear at high temperatures (above 110 K). Then we estimated the Hall coefficient $R_H$ by fitting the low field data in Fig. 3(b), and plotted the temperature-dependent $R_H$ in Fig. 3(c). The $R_H$ is negative in the whole temperature range, indicating dominant electron carriers, as well as a nonmonotonic temperature dependence, similar with that in FeS \cite{Lin}.

\begin{figure}
\centerline{\includegraphics[scale=1]{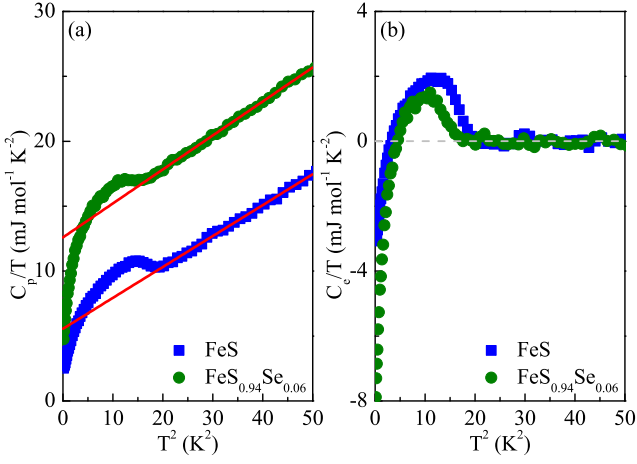}}
\caption{(a) Temperature dependence of heat capacity $C_\textrm{p}/T$ vs $T^2$ for for FeS$_{1-x}$Se$_x$ ($x$ = 0 and 0.06) single crystals. The solid curve represents the fittings using $C_\textrm{p}/T=\gamma+\beta T^2$. (b) The low temperature electronic part $C_e/T$ vs $T^2$ in zero field.}
\end{figure}

The temperature-dependent heat capacity $C_\textrm{p}/T$ in the low-temperature range as a function of $T^2$ in zero magnetic field for FeS$_{1-x}$Se$_x$ ($x$ = 0, 0.06) is depicted in Fig. 4(a). A clear specific heat jump is observed at $T_c$ for both samples, in line with the resistivity and thermopower data, indicating bulk superconductivity. The data from 4.5 to 7 K can be well fitted by $C_\textrm{p}/T = \gamma + \beta T^2$ [Fig. 4(a)], where the first term is the Sommerfeld electronic specific heat coefficient, the second term is low-temperature limit of lattice heat capacity. The derived $\gamma$ is 5.56(8) and 12.60(9) mJ mol$^{-1}$ K$^{-2}$ for FeS and FeS$_{0.94}$Se$_{0.06}$, respectively, comparable with that of FeSe. The Debye temperature $\Theta_\textrm{D} = (12\pi^4\textrm{NR}/5\beta)^{1/3}$, where $N$ = 2 is the number of atoms per formula unit and $R$ = 8.314 J mol$^{-1}$ K$^{-1}$ is the molar gas constant. From the value of $\beta$ $\sim$ 0.238(2) and 0.262(2) mJ mol$^{-1}$ K$^{-3}$, we can calculate the value of $\Theta_\textrm{D}$ is 253.7(3) and 245.7(3) K for FeS and FeS$_{0.94}$Se$_{0.06}$, respectively, larger than that of FeSe ($\sim$ 210 K) \cite{XLin}. According to the McMillan formula for electron-phonon mediated superconductivity, the electron-phonon coupling constant $\lambda$ can be deduced by
\begin{equation}
T_c=\frac{\Theta_D}{1.45}\exp[-\frac{1.04(1+\lambda)}{\lambda-\mu^{\ast}(1+0.62\lambda)}],
\end{equation}
where $\mu^{\ast}\approx$ 0.13 is the common value for Coulomb pseudo-potential \cite{McMillan}. By using the $T_c$ = 4.5(1) K and $\Theta _D$ = 253.7(3) K for FeS, we obtain $\lambda \approx$ 0.65(11), a typical value of an intermediate-coupled BCS superconductor. However, the electronic specific heat jump at $T_c$ [Fig. 4(b)], $\Delta$C$_{e}$/$\gamma T_c\approx$ 0.43, is much smaller than the weak coupling value of 1.43 \cite{Xing,McMillan}.

\begin{figure}
\centerline{\includegraphics[scale=1]{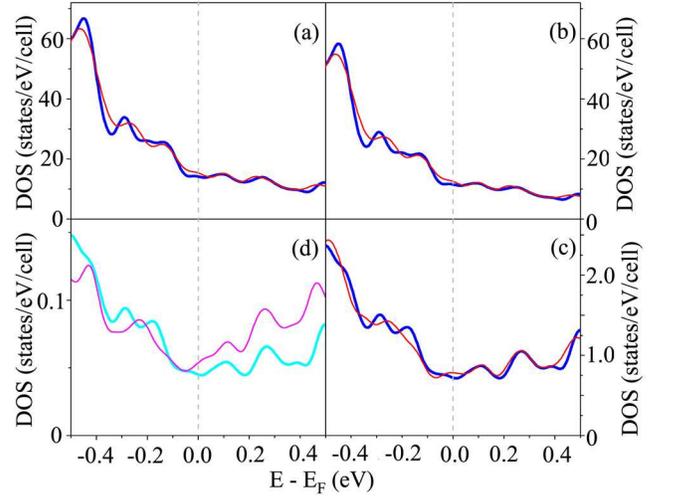}}
\caption{(Color online) The density of states in the vicinity of the Fermi level of the tetragonal Fe$_{16}$S$_{16}$ supercell (blue) and the tetragonal Fe$_{16}$S$_{15}$Se$_1$ supercell (red) for: (a) Total; (b) Total Fe part; (c) Total chalcogen part; and (d) S ion in the Fe$_{16}$S$_{16}$ supercell (cyan) and Se ion the Fe$_{16}$S$_{15}$Se$_1$ supercell (magenta).}
\label{resistivity}
\end{figure}

The electronic specific heat can be also expressed as:
\begin{equation}
\gamma = \frac{\pi^2}{2}k_\textrm{B}\frac{n}{T_\textrm{F}} = \frac{\pi^2}{3}k_\textrm{B}^2N(\varepsilon_\textrm{F}).
\end{equation}
Combining equations (3) and (5) yields: $S/T = \pm \gamma/ne$, where the units are V K$^{-1}$ for $S$, J K$^{-2}$ m$^{-3}$ for $\gamma$, and m$^{-3}$ for $n$, respectively. This relation was shown to hold in the $T$ = 0 limit for a large variety of materials, even in the presence of strong correlations, including heavy fermion metals, organic conductors, and cuprates \cite{Behnia}. Then we can define a dimensionless quantity
\begin{equation}
q=\frac{S}{T}\frac{N_\textrm{A}e}{\gamma},
\end{equation}
where $N_\textrm{A}$ is the Avogadro constant. The constant $N_\textrm{A}e$ = 9.6$\times$10$^{4}$ C mol$^{-1}$ is the Faraday constant. The $q$ gives the number of carriers per formula unit (proportional to $1/n$ where $n$ is the carrier concentration) \cite{Behnia}. The derived $q$ is $\sim$ 2.6(1) for FeS, which is almost unchanged for FeS$_{0.94}$Se$_{0.06}$ $\sim$ 2.6(2), in line with equivalent doping S at Se sites. Given the volume of unit cell $\sim$ 0.068 nm$^3$, we obtain the carrier density per volume $n \approx 5.7(3) \times 10^{21}$ cm$^{-3}$ and the Fermi momentum $k_\textrm{F} = (3\pi^2n)^{1/3} \approx 5.5(1)$ nm$^{-1}$ for FeS, which are almost unchanged with $n \approx 5.7(5) \times 10^{21}$ cm$^{-3}$ and $k_\textrm{F} \approx 5.5(2)$ nm$^{-1}$ for FeS$_{0.94}$Se$_{0.06}$, respectively. The effective mass $m^*$, derived from $k_\textrm{B}T_\textrm{F} = \hbar^2 k_\textrm{F}^2/2m^*$, increases from 4.8 $m_e$ for FeS to 9.8 $m_e$ for FeS$_{0.94}$Se$_{0.06}$, consistent with the increase of correlation strength with Se substitution. This confirms substantial increase in electron correlations in FeS$_{0.94}$Se$_{0.06}$. Furthermore, we can derive the Fermi velocity $\nu_\textrm{F}$ using $\hbar k_\textrm{F} = m^*\nu_\textrm{F}$; it decreases from 133 km s$^{-1}$ for FeS to 65 km s$^{-1}$ for FeS$_{0.94}$Se$_{0.06}$.

First-principle calculations (Fig. 5) show that the Fermi level is on a pseudo-gap-like slope of the low energy side for both compounds. The density of states (DOS) in the vicinity of Fermi level for both FeS and FeS$_{0.94}$Se$_{0.06}$ is presented in Fig. 5(a). Partial contribution of Fe and chalcogen atoms is presented in Fig. 5(b,c); the result confirms that the Fe electronic states are dominant at the Fermi level but also that there is a small increase in partial contribution of chalcogen atoms to DOS for Se-doped crystal [Fig. 4(c)] \cite{Yang2,MiaoJ}. Close inspection of the region near the Fermi level for small Se substitution on S atomic site, i.e. for FeS$_{0.94}$S$_{0.06}$ [Fig. 5(d)] indicates increase in DOS. This confirms experimental observation but also points out that DOS changes are connected with orbital states that are partially chalcogen-derived and whose contribution is not negligible \cite{Yang2}.

\section{CONCLUSIONS}

In summary, thermopower analysis of tetragonal FeS and FeS$_{0.94}$Se$_{0.06}$ indicates weak electron correlation. Both $T_c$/$T_F$ and effective mass $m$$^*$ substantially increase with slight Se substitution as small as 0.06. Since drastic changes in electronic structure and emergence of orbital-selective correlations associated with Lifshitz transition have been proposed to emerge with small lattice expansion in FeS \cite{SkornyakovSL}, our results call for investigation of electronic structure in FeS$_{1-x}$Se$_x$ crystals.

\section*{Acknowledgements}

Work at Brookhaven National Laboratory (BNL) is supported by the Office of Basic Energy Sciences, Materials Sciences and Engineering Division, U.S. Department of Energy (DOE) under Contract No. DE-SC0012704. First principle calculations were supported by the Ministry of Education, Science and Technological Development of the Republic of Serbia.\\
$^{*}$Y.L. and A.W. contributed equally to this work.\\
$^{\dag}$Present address: Los Alamos National Laboratory, Los Alamos, New Mexico 87545, USA\\
$^{\|}$Present address: College of Physics, Chongqing University, Chongqing 401331, China\\
$^{\sharp}$Present address: Material Science Division, Argonne National Laboratory, Lemont, Illinois 60439, USA\\
$^{\ddag}$yuliu@lanl.gov\\
$^{\S}$afwang@cqu.edu.cn\\
$^{\P}$petrovic@bnl.gov


\begin{references}
\bibitem{Hosono} Y. Kamihara, T. Watanabe, M. Hirano, and H. Hosono, Iron-Based Layered Superconductor La[O$_{1-x}$F$_x$]FeAs (x = 0.05-0.12) with $T_c$ = 26 K, J. Am. Chem. Soc. \textbf{130}, 3296 (2008).
\bibitem{Lai} X. Lai, H. Zhang, Y. Wang, X. Wang, X. Zhang, J. Lin, and F. Huang, Observation of superconductivity in tetragonal FeS, J. Am. Chem. Soc. \textbf{137}, 10148 (2015).
\bibitem{FeSe} F. Hsu, J. Luo, K. Yeh, T. Chen, T. Huang, P. M. Wu, Y. Lee, Y. Huang, Y. Chu, D. Yan, and M. Wu, Superconductivity in the PbO-type structure $\alpha$-FeSe, Proc. Natl. Acad. Sci. USA \textbf{105}, 14262 (2008).
\bibitem{Pachmayr} U. Pachmayr, N. Fehn and D. Johrendt, Structural transition and superconductivity in hydrothermally synthesized FeX (X = S, Se), Chem. Commun. \textbf{52}, 194 (2016).
\bibitem{Kuhn} S. J. Kuhn, M. K. Kidder, D. S. Parker, C. Cruz, M. A. McGuire, W. M. Chance, L. Li, L. Debeer-schmitt, J. Ermentrout, K. C. Littrell, M. R. Eskildsen, and A. S. Sefat, Structure and property correlations in FeS, Physica C \textbf{534}, 29 (2017).
\bibitem{Holenstein} S. Holenstein, U. Pachmayr, Z. Guguchia, S. Kamusella, R. Khasanov, A. Amato, C. Baines, H. H. Klauss, E. Morenzoni, D. Johrendt, and H. Luetkens, Coexistence of low-moment magnetism and superconductivity in tetragonal FeS and suppression of $T_c$ under pressure, Phys. Rev. B \textbf{93}, 140506(R) (2016).
\bibitem{Zhang} J. Zhang, F. Liu, T. Ying, N. Li, Y. Xu, L. He, X. Hong, Y. Yu, M. Wang, J. Shen, W. Yang, and S. Li, Observation of two superconducting domes under pressure in tetragonal FeS, npj Quantum Materials \textbf{2}, 49 (2017).
\bibitem{Man} H. Man, J. Guo, R. Zhang, R. Sch\"{o}nemann, Z. Yin, M. Fu, M. B. Stone, Q. Huang, Y. Song, W. Wang, D. J. Singh, F. Lochner, T. Hickel, I. Eremin, L. Harriger, J. W. Lynn, C. Broholm, L. Balicas, Q. Si, and P. Dai, Spin excitations and the Fermi surface of superconducting FeS, npj Quantum Mater. \textbf{2}, 14 (2017).
\bibitem{Reiss} P. Reiss, M. D. Watson, T. K. Kim, A. A. Haghighirad, D. N. Woodruff, M. Bruma, S. J. Clarke, and A. I. Coldea, Suppression of electronic correlations by chemical pressure from FeSe to FeS, Phys. Rev. B \textbf{96}, 121103(R) (2017).
\bibitem{Yang1} X. Yang, Z. Du, G. Du, Q. Gu, H. Lin, D. Fang, H. Yang, X. Zhu, and H. H. Wen, Strong-coupling superconductivity revealed by scanning tunneling microscope in tetragonal FeS, Phys. Rev. B \textbf{94}, 024521 (2016).
\bibitem{Ying} T. P. Ying, X. F. Lai, X. C. Hong, Y. Xu, L. P. He, J. Zhang, M. X. Wang, Y. J. Yu, F. Q. Huang, and S. Y. Li, Nodal superconductivity in FeS: Evidence from quasiparticle heat transport, Phys. Rev. B \textbf{94}, 100504(R) (2016).
\bibitem{Xing} J. Xing, H. Lin, Y. Li, S. Li, X. Zhu, H. Yang, and H. H. Wen, Nodal superconducting gap in tetragonal FeS, Phys. Rev. B \textbf{93}, 104520 (2016).
\bibitem{Borg} C. K. H. Borg, X. Zhou, C. Eckberg, D. J. Campbell, S. R. Saha, J. Paglione, and E. E. Rodriguez, Strong anisotropy in nearly ideal tetrahedral superconducting FeS single crystals, Phys. Rev. B \textbf{93}, 094522 (2016).
\bibitem{Terashima} T. Terashima, N. Kikugawa, H. Lin, X. Zhu, H. H. Wen, T. Nomoto, K. Suzuki, H. Ikeda, and S. Uji, Upper critical field and quantum oscillations in tetragonal superconducting FeS, Phys. Rev. B \textbf{94}, 100503(R) (2016).
\bibitem{Lin} H. Lin, Y. Li, Q. Deng, J. Xing, J. Liu, X. Zhu, H. Yang, and H. H. Wen, Multiband superconductivity and large anisotropy in FeS crystals, Phys. Rev. B \textbf{93}, 144505 (2016).
\bibitem{Yang2} Y. Yang, W. S. Wang, H. Y. Lu, Y. Y. Xiang, and Q. H. Wang, Electronic structure and $d_{x^2-y^2}$-wave superconductivity in FeS, Phys. Rev. B \textbf{93}, 104514 (2016).
\bibitem{AF1} A. Wang, L. Wu, V. N. Ivanovski, J. B. Warren, J. Tian, Y. Zhu, and C. Petrovic, Critical current density and vortex pinning in tetragonal FeS$_{1-x}$Se$_x$ (x = 0, 0.06), Phys. Rev. B \textbf{94}, 094506 (2016).
\bibitem{AF2} A. Wang and C. Petrovic, Vortex pinning and irreversibility fields in FeS$_{1-x}$Se$_x$ (x = 0, 0.06), Appl. Phys. Lett. \textbf{110}, 232601 (2017).
\bibitem{Sefat} A. S. Sefat, M. A. McGuire, B. C. Sales, R. Jin, J. Y. Howe, and D. Mandrus, Electronic correlations in the superconductor LaFeAsO$_{0.89}$F$_{0.11}$ with low carrier density, Phys. Rev. B \textbf{77}, 174503 (2008).
\bibitem{Kang} N. Kang, P. Auban-Senzier, C. R. Pasquier, Z. A. Ren, J. Yang, G. C. Chen, and Z. X. Zhao, Pressure dependence of the thermoelectric power of the iron-based high-$T_c$ superconductor SmFeAsO$_{0.85}$, New J. Phys.\textbf{11}, 025006 (2009).
\bibitem{Mun} E. D. Mun, S. L. Bud$^\prime$ko, Ni Ni, A. N. Thaler, and P. C. Canfield, Thermoelectric power and Hall coefficient measurements on Ba(Fe$_{1-x}$T$_x$)$_2$As$_2$ (T = Co and Cu), Phys. Rev. B \textbf{80}, 054517 (2009).
\bibitem{Pourret} A. Pourret, L. Malone, A. B. Antunes, C. S. Yadav, P. L. Paulose, B. Fauque, and K. Behnia, Strong correlation and low carrier density in Fe$_{1+y}$Te$_{0.6}$Se$_{0.4}$ as seen from its thermoelectric response, Phys. Rev. B \textbf{83}, 020504(R) (2011).
\bibitem{Matusiak} M. Matusiak, T. Plackowski, Z. Bukowski, N. D. Zhigadlo, and J. Karpinski, Evidence of spin-density-wave order in RFeAsO$_{1-x}$F$_x$ from measurements of thermoelectric power, Phys. Rev. B \textbf{79}, 212502 (2009).
\bibitem{Butch} N. P. Butch, S. R. Saha, X.H. Zhang, K. Kirshenbaum, R. L. Greene, and J. Paglione, Effective carrier type and field dependence of the reduced-$T_c$ superconducting state in SrFe$_{2-x}$Ni$_x$As$_2$, Phys. Rev. B \textbf{81}, 024518 (2010).
\bibitem{Kefeng1} K. Wang, H. Lei, and C. Petrovic, Thermoelectric studies of K$_x$Fe$_{2-y}$Se$_2$ indicating a weakly correlated superconductor, Phys. Rev. B \textbf{83}, 174503 (2011).
\bibitem{Kefeng2} K. Wang, H. Lei, and C. Petrovic, Evolution of correlation strength in K$_x$Fe$_{2-y}$Se$_2$ superconductor doped with S, Phys. Rev. B \textbf{84}, 054526 (2011).
\bibitem{Collignon} C. Collignon, A. Ataei, A. Gourgout, S. Badoux, M. Lizaire, A. Legros, S. Licciardello, S. Wiedmann, J. Q. Yan, J. S. Zhou, Q. Ma, B. D. Gaulin, N. Doiron-Leyraud, and L. Tillefer, Thermopower across the phase diagram of the cuprate La$_{1.6-x}$Nd$_{0.4}$Sr$_x$CuO$_4$: Signatures of the pseudogap and charge density wave phases, Phys. Rev. B \textbf{103}, 155102 (2021).
\bibitem{AndersenBM} Brian M. Andersen, Yu. S. Barash, S. Graser and P .J. Hirschfeld, Josephson effects in $d$-wave superconductor junctions with magnetic interlayers, Phys. Rev. B \textbf{77}, 054501 (2008).
\bibitem{WolfFA} F. A. Wolf, S. Graser, F. Loder, and T. Kopp, Supercurrent through Grain Boundaries of Cuprate Superconductors in the Presence of Strong Correlations, Phys. Rev. Lett. \textbf{108}, 117002 (2012).
\bibitem{YoYJ} Y. J. Jo, J. Jaroszynski, A. Yamamoto, A. Gurevich, S. C. Riggs, G. S. Boebinger, D. Larlabalestier, H. H. Wen, N. D. Zhigadlo, S. Katrych, Z. Bukowski, J. Karpinski, R. H. Liu, H. Chen, X. H. Chen and L. Balicas, High-field phase-diagram of Fe arsenide superconductors, Physica C \textbf{469}, 9 (2009).
\bibitem{HecherJ} J. Hecher, S. Ishida, D. Song, H. Ogino, A. Iyo, H. Eisaki, M. Nakajima, D. Kagerbauer and M. Eisterer, Direct observation of in-plane anisotropy of the superconducting critical current density in Ba(Fe$_{1-x}$Co$_x$)$_2$As$_2$ crystals, Phys. Rev. B \textbf{97}, 014511 (2018).
\bibitem{YuSL} Shun-Li Yu, Jing Kang and Jian-Xin Li, Band renormalization and Fermi surface reconstruction in iron-based superconductors, Phys. Rev. B \textbf{79}, 064517 (2009).
\bibitem{SkornyakovSL} S. L. Skornyakov and I. Leonov, Correlated electronic structure, orbital-dependent correlations, and Lifshitz transition in tetragonal FeS, Phys. Rev. B \textbf{100}, 235123 (2019).
\bibitem{PhysRevB.54.11169} G. Kresse and J. Furthm\"{u}ller, Efficient iterative schemes for $ab$-$initio$ total-energy calculations using a plane-wave basis set, Phys. Rev. B \textbf{54}, 11169 (1996).
\bibitem{Lee} C. H. Lee, K. Kihou, A. Iyo, H. Kito, P. M. Shirage, and H. Eisaki, Relationship between crystal structure and superconductivity in iron-based superconductors, Solid State Commun. \textbf{152}, 644 (2012).
\bibitem{Smith} R. A. Smith, \textit{Semiconductor} (Cambridge: Cambridge University Press, 1978).
\bibitem{YJSong} Y. J. Song, J. B. Hong, B. H. Min, and Y. S. Kwon, Superconducting properties of a stoichiometric FeSe compound and two anomalous features in the normal state, J. Korean Phys. Soc. \textbf{59}, 312 (2011).
\bibitem{Caglieris} F. Caglieris, F. Ricci, G. Lamura, A. Martinelli, A. Palenzona, I. Pallecchi, A. Sala, G. Profeta, and M. Putti, Theoretical and experimental investigation of magnetotransport in iron chalcogenides, Sci. Technol. Adv. Mater. \textbf{13}, 054402 (2012).
\bibitem{Pallecchi} I. Pallecchi, F. Caglieris, and M. Putti, Thermoelectric properties of iron-based superconductors and parent compounds, Supercond. Sci. Technol. \textbf{29}, 073002 (2016).
\bibitem{Subedi} A. Subedi, L. Zhang, D. J. Singh, and M. H. Du, Density functional study of FeS, FeSe, and FeTe: Electronic structure, magnetism, phonons, and superconductivity, Phys. Rev. B \textbf{78}, 134514 (2008).
\bibitem{Barnard} R. D. Barnard, \textit{Thermoelectricity in Metals and Alloys} (Taylor \& Francis, London, 1972).
\bibitem{Cohn} J. L. Cohn, S. A. Wolf, V. Selvamanickam, and K. Salama, Thermoelectric power of YBa$_2$Cu$_3$O$_{7-\delta}$: Phonon drag and multiband conduction,Phys. Rev. Lett. \textbf{66}, 1098 (1991).
\bibitem{Behnia} K. Behnia, D. Jaccard and J. Flouquet, On the thermoelectricity of correlated electrons in the zero-temperature limit, J. Phys.: Condens. Matter. \textbf{16}, 5187 (2004).
\bibitem{Miyake} K. Miyake and H. Kohno, Theory of Quasi-Universal Ratio of Seebeck Coefficient to Specific Heat in Zero-Temperature Limit in Correlated Metals, J. Phys. Soc. Jpn. \textbf{74}, 254 (2005).
\bibitem{Sarrao} J. L. Sarrao and J. D. Thompson, Superconductivity in Cerium- and Plutonium-Based `115' Materials, J. Phys. Soc. Jpn. \textbf{76}, 051013 (2007).
\bibitem{Moriya} T. Moriya and K. Uedo, Antiferromagnetic spinfluctuation and superconductivity, Rep. Prog. Phys. \textbf{66}, 1299 (2003).
\bibitem{XLin} J. Y. Lin, Y. S. Hsieh, D. A. Chareev, A. N. Vasiliev, Y. Parsons, and H. D. Yang, Coexistence of isotropic and extended $s$-wave order parameters in FeSe as revealed by low-temperature specific heat, Phys. Rev. B \textbf{84}, 220507(R) (2011).
\bibitem{McMillan} W. L. McMillan, Transition Temperature of Strong-Coupled Superconductors, Phys. Rev. \textbf{167}, 331 (1968).
\bibitem{MiaoJ} J. Miao, X. H. Niu, D. F. Xu, Q. Yao, Q. Y. Chen, T. P. Ying, S. Y. Li, Y. F. Fang, J. C. Zhang, S. Ideta, K. Tanaka, B. P. Xie, D. L. Feng, and Fei Chen, Electronic structure of FeS, Phys. Rev. B \textbf{95}, 205127 (2017).

\end{references}
\end{document}